\documentclass[twocolumn,aps,prb,showpacs,amsmath,amssymb,psfrac]{revtex4}

\usepackage{amsmath}

\usepackage{epsfig}
\usepackage{graphicx}
\usepackage{dcolumn}
\usepackage{bm}
\usepackage{psfrag}

\def\gl{\lower.35em\hbox{$\stackrel{\textstyle>}{\textstyle<}$}}
\def\gapp{\lower.35em\hbox{$\stackrel{\textstyle>}{\sim}$}}
\def\lapp{\lower.35em\hbox{$\stackrel{\textstyle<}{\sim}$}}

\newcommand{\vf}{v_{\rm F}}

\newcommand{\ef}{\epsilon_{\rm F}}

\begin{document}

\title{Electron-electron interactions and charging effects in graphene quantum
  dots}
\author{B. Wunsch$^{1,2}$, T. Stauber$^{2}$, and F. Guinea$^{2}$}
\affiliation{$^1$ Departamento de F\'isica de Materiales,
  Facultad de Ciencias F\'isicas,  Universidad Complutense de Madrid, E-28040
  Madrid, Spain.\\$^2$ Instituto de Ciencia de Materiales de Madrid, CSIC, Cantoblanco,
  E-28049 Madrid, Spain.}
\date{\today}
\begin{abstract}
  We analyze charging effects in graphene quantum dots. Using a simple
  model, we show that, when the Fermi level is far from the neutrality
  point, charging effects lead to a shift in the electrostatic
  potential and the dot shows standard Coulomb blockade features. Near
  the neutrality point, surface states are partially occupied and the
  Coulomb interaction leads to a strongly correlated ground state
  which can be approximated by either a Wigner crystal or a Laughlin
  like wave function.  The existence of strong correlations modify the
  transport properties which show non equilibrium effects, similar to
  those predicted for tunneling into other strongly correlated
  systems.

\end{abstract}
\pacs{73.63.Kv, 73.23.Hk, 73.43.Lp}
%
%
\maketitle
\section{Introduction}
Graphene has attracted a great deal of attention, because of its novel
fundamental properties and its potential applications\cite{GN07}. The interest
on graphene devices has motivated recent research on the transport properties
of small devices\cite{HSSTYG07,HOZK07,WDM07,OHLMV07}. Features such as
charging effects and quantum confinement are of crucial importance for their
understanding\cite{MJFP06,SGN07b}. The confinement of electrons, and the
observation of Coulomb blockade effects has already been demonstrated
experimentally\cite{Betal05,GN07}. Note that the confinement of electrons in
graphene is not trivial, due to the {\it Klein's paradox}\cite{KNG06}, which
makes potential barriers transparent for normally incident quasi-particles.
Electrons in graphene can be confined, however, by exploiting the angular
dependence of scattering at a barrier\cite{SE07}.

For graphene layers, electron-electron interaction is usually neglected, including works
on localization\cite{Aleiner06} even though disorder enhances the effect of
interaction.\cite{Stauber05} The reasoning for this is to assume a ``normal''
ground-state at zero doping - characterized by a semi-metal. Because the
kinetic and interaction energy equally scale with the carrier
density, the interaction does not become important at finite doping, either.
It is thus well agreed on that at finite doping electron-electron interaction can
be treated within the random-phase approximation
(RPA).\cite{Wunsch06,Hwang07} Nevertheless, at the Dirac point RPA seems to
fail leading to a novel plasmon mode in graphene.\cite{Mishchenko07} Also in a
quantum dot, we find that electron-electron interaction has to be treated
differently for doping regimes close to and away from the Dirac point.

The main part of this work is the prediction and characterization of strongly
correlated few-electron states in graphene quantum dots. Similar studies have
been performed previously for semiconducting quantum
dots\cite{Baranger07,YL02,YL04,Reimann}. In order to obtain strongly
correlated ground states the Coulomb interaction has to dominate over the
other energy scale, namely the shell structure of the single particle spectrum
determined by the confinement. This is typically achieved by either using
strong magnetic fields\cite{YL02,YL04}, so that the single particle levels
form highly degenerate Landau levels, or using rather weak
confinement.\cite{Baranger07} Interestingly, strongly correlated states
naturally arise already in small graphene quantum dots even without magnetic
field.  The reason for that is the appearance of a highly degenerate zero
energy band of surface states, which is strongly affected by Coulomb
interaction.  Close to half filling these states are occupied by few electrons
which are strongly correlated and can be approximated by a Laughlin like wave
function or alternatively by a quasi one-dimensional Wigner crystal.

The paper is organized as follows. In section II, we present a simple model which
allows us to describe qualitatively the charging of a graphene dot. In
section~III, we show that the charging properties of the graphene dot are in
agreement with Coulomb blockade theory when the Fermi energy is far from the
neutrality point. Thereafter, we show in section~IV that close to the
neutrality point charging effects are strongly modified by the presence of
midgap states, associated to the edges\cite{FWNK96}.  We show that electrons
occupying these midgap states form a strongly correlated state which is
characterized in detail. In section~V, we then discuss implications for
transport properties. We close with conclusions and outlook.

{
\section{The model}
The linearized tight-binding Hamiltonian for a graphene
sheet with circular symmetry is given by
\begin{align}
H_s= \vf \begin{pmatrix}
0&e^{is\theta}(-is\partial_r+\frac{1}{r}\partial_\theta)\\
e^{-is\theta}(-is\partial_r-\frac{1}{r}\partial_\theta)&0
\end{pmatrix}\;,
\label{kinetic}
\end{align}
where $s=\pm$ determines the valley. We assume that the dot is
ballistic, i.e., with no internal disorder. The general solutions with
energy $\epsilon_k = \pm \vf k$ are of the type
\begin{equation}
\left( \begin{array}{c} \Psi_s^A ( r , \theta ) \\ \Psi_s^B ( r , \theta )
  \end{array} \right) = \left( \begin{array}{c} J_{m+s} ( k r ) e^{i (
    m+s) \theta} \\ \mp i J_m ( kr ) e^{i m \theta} \end{array} \right)
\label{wavefunction}
\end{equation}
with $J_m ( x )$ denoting the $m$-th Bessel function.  The dot has a circular
shape with radius $R$. The circular symmetry of the dot allows us to classify
the solutions according to their angular momenta.

In order to analyze the
possible role of surface states, we assume that the boundary conditions at the
edges are those appropriate for a zig-zag graphene edge ending always on the
same lattice site\cite{FWNK96}
\begin{equation}
\Psi_s^A ( R , \theta ) = 0\;.
\label{bc}
\end{equation}
The boundary condition is not experimentally realizable for a circular dot, though
it enables a detailed analysis of the interplay between
Coulomb interaction and surface states. For the chosen boundary
condition the wave vector is quantized by $k=z_{n m}/R$, where $z_{nm}$
denotes the $n$-th root of the $m$-th Bessel function, $J_m(z_{nm})=0$.
In addition to the finite energy states given in
Eq.~(\ref{wavefunction}) the boundary condition allows for surface
states which can be written as
\begin{equation}
\left( \begin{array}{c} \Psi_s^A ( r , \theta ) \\ \Psi_s^B ( r , \theta )
  \end{array} \right) = \left( \begin{array}{c}0\\ \sqrt{\frac{m+1}{\pi R^{2
        ( m+1)}}} r^m e^{i s m \theta} \end{array} \right)\;,
\label{edge_state}
\end{equation}
with $m \geq 0$ to guarantee normalizability. Note, that for the
surface states the angular momentum is given by $s m$, see
Eq. (\ref{edge_state}) and that these functions have an analytical
dependence on either $z = x + i y$ or $\bar{z} = x - i y$. Discrete
lattice effects impose a maximal (absolute) value on the angular
momentum of order $m_{max} \sim R/a$, where $R$ is the radius of the
disk and $a$ is a length comparable to the lattice spacing.

Charging effects arise from electron-electron interaction which is generally described by
\begin{equation}
H_C=\frac{e^2}{4\pi
  \epsilon_0\epsilon}\sum_{n<n'}\frac{1}{|\bm r_n-\bm r_{n'}|} \label{coulomb}\;.
\end{equation}
The total Hamiltonian is given by the sum of Eqs. (\ref{kinetic}) and
(\ref{coulomb}). We note that both parts scale as $1/R$.  Furthermore in
graphene $(e^2 /4\pi \epsilon_0\epsilon)/ \hbar v_F\approx 1$, so that both
single particle and interaction
energies can be expressed in units of $(e^2 /4\pi
\epsilon_0\epsilon R)=\hbar v_F/R$, as will be done throughout this paper.

Charging effects are mostly determined by the overall geometry of the dot, so
that the lack of disorder in the model described here does not change
qualitatively the main features of Coulomb blockade. Graphene dots have, most
likely, rough edges. Hence, the possible surface states are confined to
certain regions of the edges. The model overestimates the number of surface
states of a given dot. On the other hand, wave functions localized in the
angular coordinate $\theta$ can be built from the wave functions in
Eq.(\ref{wavefunction}) or Eq.(\ref{edge_state}). A dot where the edge has a
region of size $l$ of the zig-zag type has states localized at the edge with
an angular width $\Delta \theta \sim l / R$. These states will be
approximately described by superpositions of states with angular momenta $m
\lesssim R / l$. Hence, when $R / l \gg 1$ and $l / a \gg 1$, these states,
which will change over distances larger than the lattice spacing, will be well
described by superpositions of the states derived from our continuum model.

\section{Charging effects away from the Dirac energy}
We analyze the effects induced by increasing the number of electrons in the
dot using the Hartree approximation. The self-consistent Hartree potential
describes, within a mean field approximation, the screening of charges within
the dot. We assume that a half filled dot is neutral, as the ionic charge
compensates the electronic charge in the filled valence band. Away from half
filling, the dot is charged. Then, an electrostatic potential is induced in
its interior, and there is an inhomogeneous distribution of
charge\cite{charge}. We describe charged dots by fixing the chemical
potential, and obtaining a self-consistent solution where all electronic
states with lower energies are filled.  The Hartree approximation should give
a reasonable description when Coulomb blockade effects can be described as a
rigid shift of the electrostatic potential within the dot\cite{AL91,SET}.

The Hartree potential needs to be calculated self consistently which must be
done numerically, despite of the simplicity of the model. The Dirac equation
for each angular momentum channel is discretized, and an effective tight
binding model is defined for each channel. Details are given in Appendix
\ref{App:TightBinding}.

\begin{figure}[t]
  \begin{center}
    \includegraphics*[width=6cm,angle=-90]{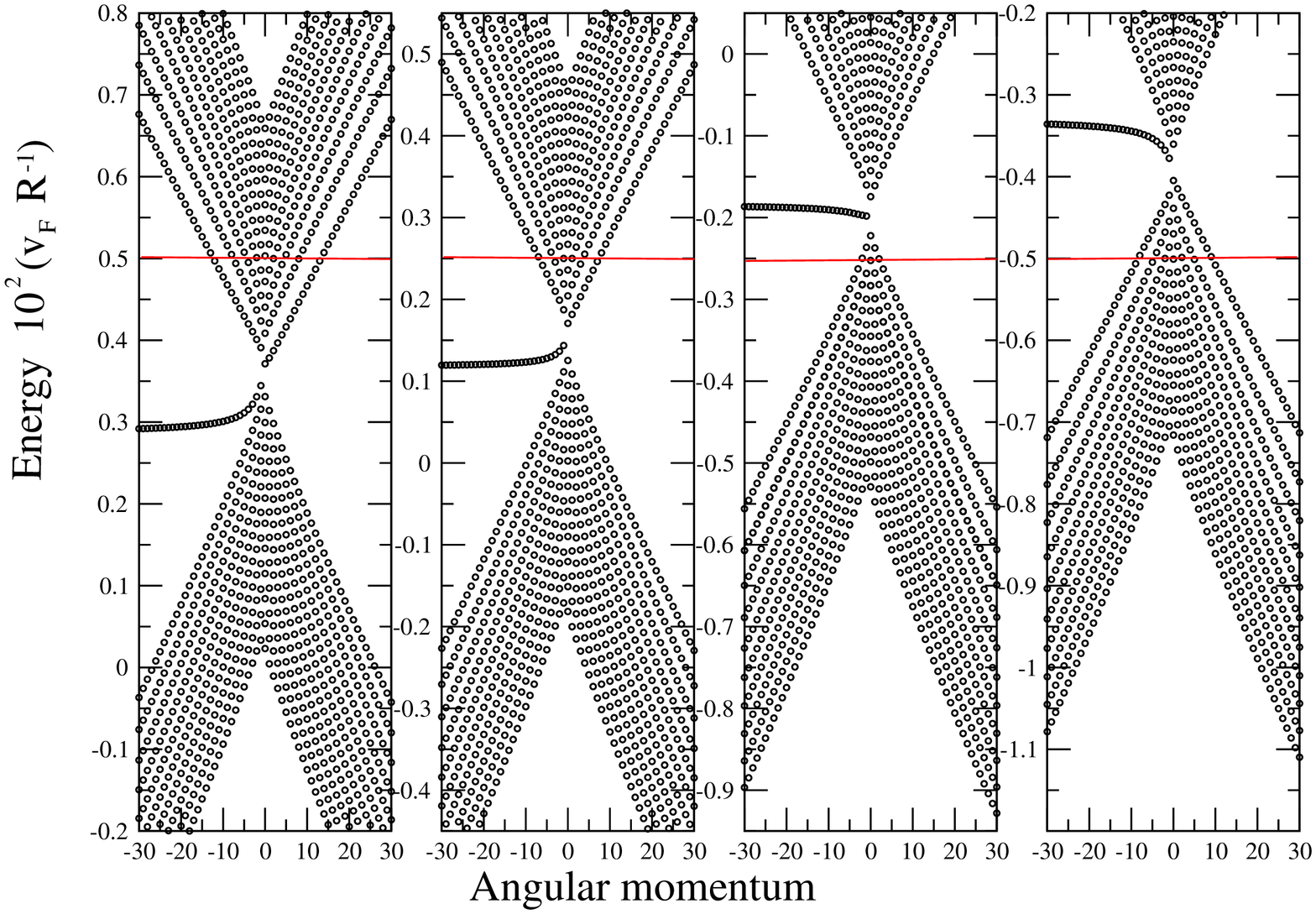} \\
    \includegraphics*[width=6cm,angle=-90]{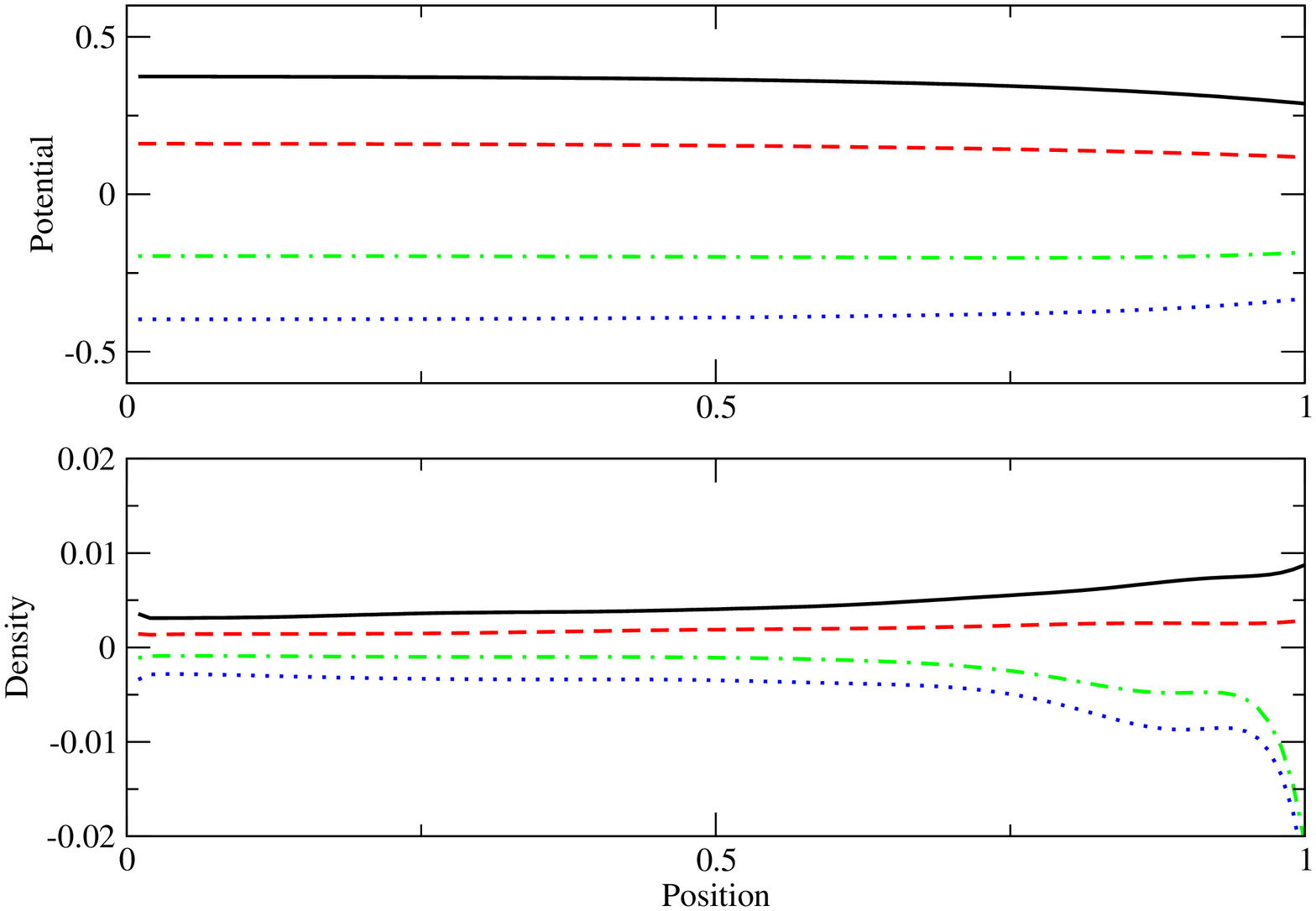}
    \caption{(Color online). Electronic structure of a quantum dot in
    the Hartree approximation. All energies are in units of $\hbar \vf N/ R$
    with $N=100$ and the position in units of $r / R$.  \\ 
    Top part: Electron energies as function of the
    angular momentum, for one valley in the Brillouin Zone, and
    different values of the chemical potential. Left: $\ef =
    0.5$. Center left: $\ef = 0.25$. Center right: $\ef =
    -0.25$. Right: $\ef = -0.5$. The total number of states per valley
    is 12200. The number of occupied states per valley is 6155 (left),
    6122 (center left), 6066 (center right), and 6038, right. Bottom
    part: Hartree potential (top), and charge density (bottom) for the
    four values of the chemical potential considered in the top part:
    Black (solid line), $\ef = 0.5$. Red (dashed line), $\ef =
    0.25$. Green (dash-dotted line), $\ef = -0.25$. Blue (dotted
    line), $\ef = -0.5$. }
\label{fig0}
\end{center}
\end{figure}
The conservation of the angular momentum allows for the possibility of solving
dots with a large number of electrons. Typical results for dots charged with
electrons or holes away from the Dirac energy are shown in Fig.~\ref{fig0}.
The calculation has been done in a discrete lattice with $N=100$ sites (see
the Appendix). The Hartree potential changes little within the dot and, to a
first approximation, the deviation from neutrality of the dot can be
approximated by a rigid shift of the electrostatic potential\cite{charge}.

\section{Charging effects near the neutrality point}
The Hartree calculations mentioned above fail to give a self consistent
solution when the surface band is partially occupied and a more advanced
treatment of the interaction has to be applied. This also implies that
deviations from conventional Coulomb blockade can be expected in this regime.

Instead of treating the interactions within a mean field approach we therefore
employ the method of configuration interaction to fully take into account all
correlations within the truncated Hilbert space of surface states. The
truncation of the Hilbert space can be justified by the energy gap to extended
states of finite energy which in our model is given by $2.4 \hbar v_F/R$. In
principle, the effect of the extended states can be added to the following
analysis as a perturbation, but we do not expect qualitative changes of our
main conclusions.

In the following we deal, therefore, with a few-electron problem and consider
the eigenspectrum of $N$ interacting electrons occupying surface states.  The
interaction is described in Eq.~(\ref{coulomb}).  Since the screening of
electron-electron interaction is known to be poor close to half filling, it
seems sensible to consider a long-ranged interaction rather than a
(point-like) Hubbard interaction.  In addition to the particle number, the
few-electron wave function can be characterized by the valley-polarization
$I_z=\sum_n s_n$ (in absence of inter-valley scattering), the total angular
momentum $M=\sum_{n}s_n m_n$ as well as by the quantum number $S^2,S_z$ for
the total spin.  In the following we limit ourselves to valley and spin
polarized solutions. While spin polarized electrons cannot interact with each
other via a point-like Hubbard interaction (due to Pauli principle), the long
ranged Coulomb interaction will give rise to highly correlated spin-polarized
states as shown in the rest of the paper.

Electron-electron interaction tries to maximize the distance between the
electrons, which leads to a correlated ground state. The few-electrons ground
state for $M\to\infty$ is given by a classical Wigner crystal where the $N$
electrons are localized at $r=R$ and $\theta_n=2\pi n/N$, thus minimizing the
Coulomb energy, see insets in Fig.~\ref{fig1}. It is important to note that
due to the localization, the truncation of the Hilbert-space to include only
surface states is still (in fact, even better) justified in the presence of
electron-electron interactions.

Surface states are characterized by only populating one of the two
sub-lattices and thus avoiding the kinetic energy due to nearest-neighbor
hopping $t$. However, next-nearest neighbor hopping $t'\approx
t/10\approx0.3$eV connects sites within the same sub-lattice so that surface
states gain some finite kinetic energy and the zero energy band becomes
dispersive. This kinetic term delocalizes the wave-function of the surface
states and leads to a stable few-electron ground-state with finite angular
momentum $M_{0}$. From Ref.\cite{PGN06}, the kinetic energy due to next-nearest
neighbor hopping reads $t'a^2 p^2$, with $a$ the lattice spacing and $p$ the
momentum operator.  As shown in appendix \ref{App:KineticEnergy}  the
Hamiltonian for next nearest neighbor hopping $H_{kin}$ can be written to lowest
order perturbation in $t'$ as
\begin{eqnarray*}
H_{kin}=\frac{\hbar
  v_F}{R}\frac{3 a}{10 R}\sum_m   m (m+1) c_m^\dag c_m\;.
\label{Hkin}
\end{eqnarray*}
This kinetic term competes with the Coulomb interaction, since it
reduces the Coulomb correlations of the ground state.  This
competition is also visible in the dependence of the few-electron
energy on the total angular momentum as shown in Fig.~\ref{fig1}. In
the absence of next-nearest neighbor hopping (solid line) the energy
decreases with increasing angular momentum (except for oscillations
discussed below), since states of higher angular momentum have lower
Coulomb energy. However, when next-nearest neighbor hopping is
included then the occupation of states with large angular momentum is
hindered and the energy as function of angular momentum shows a well
defined minimum.  We note that the ratio between kinetic energy and
Coulomb energy increases with decreasing dot size (the numerical
calculations are done for $R=22$nm). Consequently for smaller dots the
angular momentum of the ground-state decreases. In the studied
subspace of valley and spin polarized electrons, the minimal angular
momentum is given by $M_{\rm min}=N(N-1)/2$.


\subsection{Trial functions for the correlated ground-state}

The lack of well converged Hartree solutions, which are given by Slater
determinants, imply that the wave-function which describes the surface states
in the presence of charging effects is strongly correlated. We have chosen two
ans\"atze which are compared to the numerically exact solution.

\subsubsection{Laughlin wave function}
The appearance of a partially filled degenerate energy band separated by an energy
gap from the rest of the spectrum strongly resembles Fractional
Quantum Hall physics. However, now the zero energy band is caused by
the boundary condition and the gap is due to the confinement rather
than due to high magnetic fields.  Not only the band structure is
similar in both systems, but also the form of the one-particle
states is similar. Both the surface states as well as the orbitals
of the lowest Landau level (in symmetric gauge) depend on $z^m$.

This analogy can be used to propose a trial wave function much like Laughlin's
original wave function for the ground-state in the Fractional Quantum Hall
regime\cite{L83}
\begin{equation}
\psi(z_1,z_2,\dots,z_N)={\mathcal C} \prod_{i<j}(z_i-z_j)^p\;,
\label{laughlin}
\end{equation}
with $p$ odd to ensure antisymmetry and $\mathcal C$ a normalization constant.
These wave functions have a well defined total angular momentum, $M = p N (N
-1 )/ 2$. For $p=1$, this is the minimal possible angular momentum of
$N$-fully polarized electrons occupying surface states and the trial wave
function (which in this case is given by a single Slater determinant) is the
exact eigenstate. With increasing value of $p = 1, 3 , 5 , \cdots$ the
correlations increase and the wave function is given by an increasing number
of superposed Slater determinants, much like Laughlin's original wave function
for the Fractional Quantum Hall state\cite{L83}. The Laughlin like
wavefunction in Eq.~(\ref{laughlin}) is a parameter free trial wavefunction
that conserves the present symmetries (i.e. total angular momentum) and that
can be uniquely expressed in the subspace of surface states. Furthermore the
factors $(z_i-z_j)^p$ create extended holes around each electron, which
minimizes the Coulomb energy and explains the good agreement between the trial
wave function and the numerically calculated ground state. We note however, that
we use the similarity between the studied system and the Fractional Quantum
Hall effect only to get a trial function for the ground state, while we do not
analyze the similarities between both systems in the excitation spectrum.

\subsubsection{Wigner crystal}
An obvious alternative to the FQHE like wave function described above is that
of a Wigner crystal.  The surface states are maximal at the border of the dot
and the system resembles a one-dimensional system. In order to minimize the
Coulomb energy, it is therefore favorable
to superpose the wave functions in such a way that electrons are maximally
separated in angle.  We write such a trial function as
\begin{align}
 | \Psi_{WC} \rangle &\equiv \hspace{-6ex}\sum_{\tiny \begin{array}{c} m_1 , m_2 , \cdots
   ,m_N \\
   m_1 + m_2 + \cdots
   + m_N = M \end{array}}\hspace{-6ex} e^{-i \sum_n \frac{2 \pi n  m_n}{ N}} \mathop \prod_{m_n}
   (m_n+1)^w c^\dag_{m_n} | 0 \rangle
\label{wc}
\end{align}
where the operator $c^\dag_{m_n}$ creates a state with momentum $m_n$.
Note that due to the constraint imposed by the antisymmetry
requirement the wave function can only be defined for total angular
momenta of the form $M=N(N-1)/2+j N$ where $j$ is a positive integer.
While the phase factor guarantees for the angular correlations in the
wave function, we use the factor $(m_n+1)^w$ to optimize radial
correlations (note that the normalization constant of a surface state
is proportional to $\sqrt{m+1}$). For a strictly one
dimensional system the usual definition of a quasi-classical Wigner
crystal implies $w=0$. In the numerical calculations, we choose $w$
such that the wave function optimizes the ground-state energy
($w\approx2$ for low $M$ and $t'$).

\subsection{Correlations}
For $M\rightarrow\infty$ and without next-nearest neighbor hopping, the
quantum mechanical configuration approaches the classical one, which minimizes the
Coulomb energy by pinning the electrons at $r=R$ and $\varphi_n=2\pi
n/N$. However, due to the rotational symmetry of our problem, the ground-state
is a superposition of all orientations such that there is a constant density
distribution around the circumference. To characterize a Wigner crystal or
more generally a density correlated system, one thus has to look at the
density-density correlation function
\begin{align}
\label{DenesityDensityCorrelation}
C_M^N(\bm r_0,\bm r)=\langle N,M;0|\frac{\sum_{i\neq j} \delta(\bm r_0-r_i),\delta(\bm r-r_j)}{N(N-1)}|N,M;0\rangle
\end{align}
where $|N,M;0\rangle$ is the ground state of the $N$ particle system to fixed
angular momentum $M$ and $i,j \in {1,..,N}$.

\subsection{Effect of Disorder}
We can use the same truncated basis to study disorder due to the roughness of
the edges. Due to the flat dispersion in the absence of disorder, single
particle states tend to localize near imperfections of the edge, however one
can show that the degeneracy of the zero energy states is only reduced by the
number of impurities, which can be assumed to be much smaller than the number
of surface states. The correlated state found in the presence of interactions
will also be pinned by disorder, leading to glassy features\cite{ES75}.

\subsection{Numerical results}
\begin{figure}[t]
  \begin{center}
    \includegraphics*[width=0.8\linewidth,angle=0]{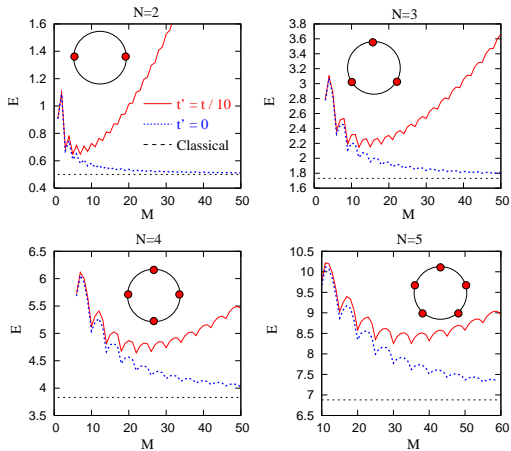}
    \caption{(color online). Ground-state energy (in units of
        $\frac{e^2}{4 \pi \epsilon_0 \epsilon}\frac{1}{R}$) for $N=2,3,4,5$
        electrons occupying surface states as function of the total
        angular momentum $M$ with (red solid line) and without (blue
        dotted line) next-nearest neighbor hopping $t'$ (assuming
        $R=22$nm). For $t'=0$ and $M\to \infty$ the energy approaches
        the classical energy of $N$ point charges on a disc. The
        classical configuration is shown in the inset and the
        classical energy is indicated by the constant dashed line.}
\label{fig1}
\end{center}
\end{figure}
\begin{figure}[t]
  \begin{center}
    \includegraphics*[width=\linewidth,angle=0]{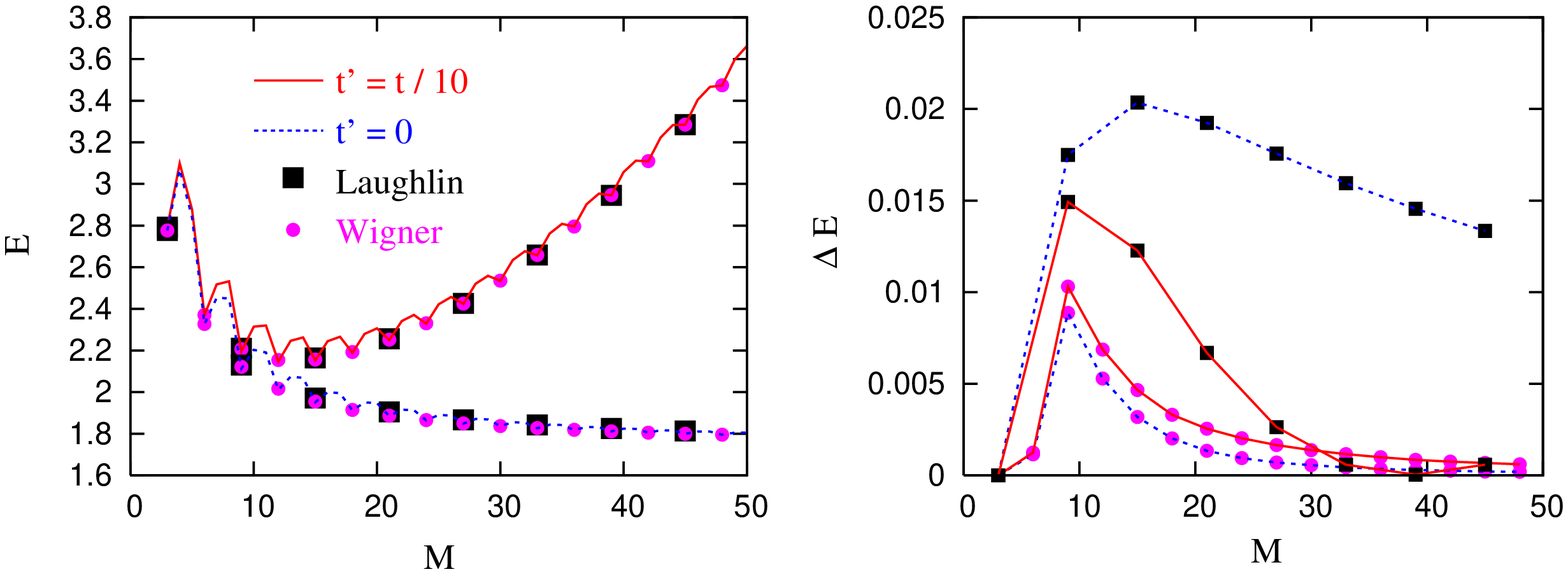}
    \caption{(color online). Ground-state energy for $N=3$ electrons with
      (full) and without (dotted) next-nearest neighbor hopping $t'$ in
      comparison with the one obtained from the trail functions. The right
      hand side shows energy differences. Energies are in units of
      $\frac{e^2}{4\pi \epsilon_0 \epsilon}\frac{1}{R}$.}
\label{figComp}
\end{center}
\end{figure}

Fig.~\ref{fig1} shows the energy of the lowest lying spin- and
valley-polarized eigenstate of each total angular momentum $M$ for
$N=2,3,4,5$ electrons occupying surface states.  The energies are
obtained by numerically diagonalizing the few-particle Hamiltonian in
this subspace.  The dotted line in Fig.~\ref{fig1} shows the results
if next nearest neighbor interaction is neglected, so that the total
Hamiltonian consists of the Coulomb interaction, only. We note two
main features in that case. First, the energy oscillates as function
of the angular momentum with local minima at $M=M_{min}(N)+j N$, where
$M_{min}(N)=N(N-1)/2$ denotes the minimal angular momentum of $N$- spin
and valley-polarized electrons and $j$ is a positive integer. Only at
these angular momenta the angular correlations between the electrons
can be fully developed. This can be seen in the correlation functions
discussed below and in the fact that only for these distinct angular
momenta a Wigner trial function can be constructed.  The second
feature visible in Fig.~\ref{fig1} is that the energy generally
decreases with increasing angular momentum for $t'=0$ and it finally reaches
the classical limit corresponding to $N$ point charges on the dot. The classical
configurations are shown in the insets, and their energies are 
indicated by the constant dashed lines.

For finite $t'$ (see solid line in Fig.~\ref{fig1}), the Hamiltonian
is supplemented by a kinetic term given in Eq.~(\ref{Hkin}) that
competes with the Coulomb interaction.  Since the kinetic energy of
surface states increases quadratically with their angular momentum the
cost in kinetic energy exceeds for large total angular momenta the
gain in Coulomb energy connected with an increase in angular
momentum. Thus the $N$-electron system now has a ground-state with
well defined angular momentum $M_0$. The ratio between the kinetic
term and the Coulomb energy grows for decreasing dot sizes, which also
leads to a decrease in $M_0$.

In Fig.~\ref{figComp} we compare the numerically obtained energies for $N=3$
electrons with that of the two trial functions described above. The data for
the Laughlin-like wave function (defined in Eq.~(\ref{laughlin})) is indicated
by squares while the data for a Wigner-crystal-like wave function (defined in
Eq.~(\ref{wc})) is labeled by filled circles. First, we note that the energies
of both trial wave functions differ by less than $1\% $ from the numerical data.
As noted above the Wigner-crystal-like wave function can be constructed
for each angular momentum where the few-electron energy shows a minimum. In
contrast a Laughlin-like wave function only exists for each $N-1$-th minimum.
It is interesting to note that the Laughlin-like wave function becomes better
for finite $t'$ than for $t'=0$.

In Fig.~\ref{figComp} we optimized the free parameter $w$ in the
Wigner wave function for each $M$ separately, which leads to this extremely
good agreement with the exact data for both zero and finite $t'$.
We note, however, that the optimal value was $w \approx 2$ for all $M$ in the
case of $t'=0$, while we strongly increased $w$ with increasing $M$ for finite $t'$.

Fig.~\ref{figCorr} shows the density plot of the exact, symmetrized
density-density correlation function $\widetilde{C}_M^N(\bm r)=\sum_{i=1}^N C_M^N(R,i
2\pi/N;\bm r)$ for $N=2,3,4,5$ electrons and for $M=N+M_{\rm min}$.  The
$N$-fold symmetry, which is typical of a 1-d Wigner crystal is clearly seen. We
note that also the trail wave function show these correlations, which
explains the good agreement of its energies with the exact one.

In Fig.~\ref{figCorr3}, the angular correlations along the perimeter
of the dot is shown for $N=3$. An electron is fixed at $\theta=0$ and
$r=R$ and the probability of finding another electron at a given angle
is plotted.  The left hand side of Fig.~\ref{figCorr3} illustrates
that correlations are maximally developed at the distinguished angular
momentum $M=jN+M_{min}$ (here $j=3$), while for other angular momenta
the correlations are washed out. On the right hand side of
Fig.~\ref{figCorr3}, we see that the density-density correlations are
more pronounced for higher angular momentum (here $j=15$) while the
kinetic energy $t'$ reduces these correlations, which again is a
manifestation of the competition between Coulomb interaction and
next-nearest neighbor hopping.

\begin{figure}[t]
  \begin{center}
    \includegraphics*[width=0.8 \linewidth,angle=0]{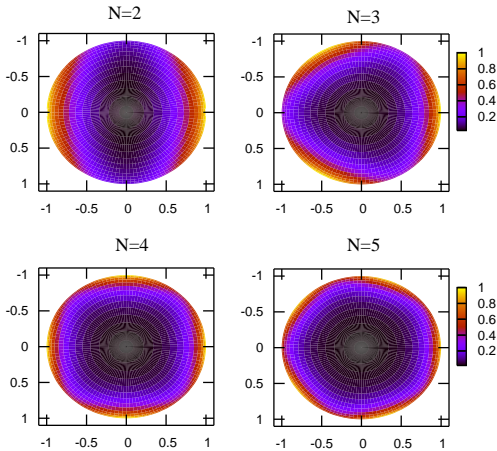}
    \caption{(color online) Density plot of the exact, symmetrized
      density-density correlation function $\widetilde{C}_M^N(\bm r)$ for
      $N=2,3,4,5$ particles and total angular momentum $M=N+M_{\rm
      min}$.}
\label{figCorr}
\end{center}
\end{figure}

\begin{figure}[t]
  \psfrag{C}{\tiny \bf $\hspace{-4ex}C_{M}^{3}(R,0;R,\theta)$}
  \begin{center}
    \includegraphics*[width=\linewidth,angle=0]{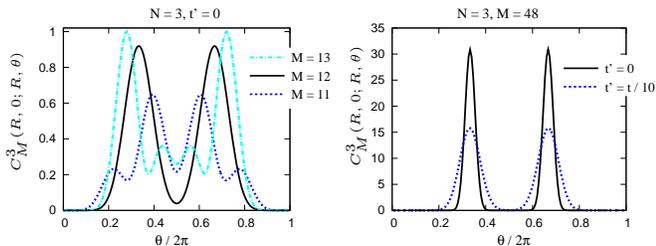}
    \caption{(color online) Angular correlations $C_{M}^{N}(R,0;R,\theta)$ for
      $N=3$ along the perimeter of the dot for various total angular momentum
      $M$ (left hand side) and various next-nearest neighbor hopping $t'$
      (right hand side). }
\label{figCorr3}
\end{center}
\end{figure}
%
%

\section{Transport properties}
The addition of one electron to the dot, in the regime where the surface
states are partially occupied, not only charges the dot and shifts the
electrostatic potential, but changes the correlated wave function as well.
Hence, one expects a correction to the local density of states in the dot,
which is energy dependent, in a similar way to Anderson's orthogonality
catastrophe\cite{A67}, or the singularity in the X-ray core level
photoemission\cite{ND69,M93}. Such Fermi edge singularities have also been
discussed in relation to transport in quantum dots and
nanotubes\cite{UG91,BHGS00,AL04,G05}.

The correlated state which describes the surface states of the graphene quantum
dot resembles a one dimensional system localized along the surface. In this
respect, the tunneling into this state can also be analyzed within the
related framework of tunneling into correlated one dimensional
metals\cite{KF92}. In this case, and in those describe before, one expects
that the tunneling density of states of the dot will be described by a
power law. We have computed numerically the spectral function

\begin{align}
A_N(\omega)\propto\sum_{M,n} & \Big|\langle N-1,M_0;0|\sum_{m=0}^{m_{\rm max}}
c_m|N,M;n\rangle\Big|^2 \notag\\
&\times\delta(E_n^{N,M}-E_0^{N-1,M_0}-\omega)
\label{SpectralFunction}
\end{align}
where $c_m$ annihilates a particle with angular momentum $m$.  Next nearest
neighbor hopping $t'$ causes a finite total angular momentum $M_0$ of the
$N-1$-electron ground state and due to momentum conservation the angular
momentum of the N-electron state is given by $M=M_0+m$. We restrict the
$m$-summation by an upper angular momentum.

Results for the spectral function are shown in
Fig.~\ref{fig4} which are characterized by a sharp peak, reminiscent to the
delta peak of the non-interacting system.  Due to the electron-electron
interaction, this peak is smeared out and decays as a power law decay, in
qualitative agreement with the arguments mentioned above.  There is a clear
convergence for low energies as function of the maximal angular momentum.

\begin{figure}[t]
  \begin{center}
    \includegraphics*[width=0.8\linewidth,angle=0]{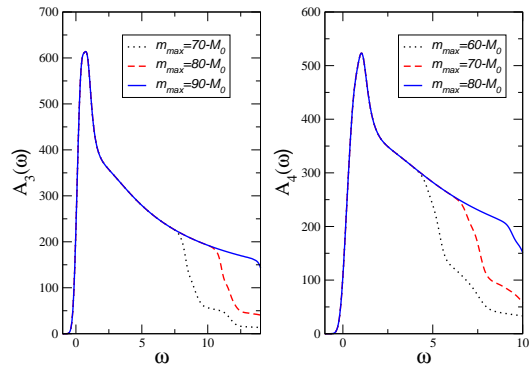}
    \caption{(color online). The (unnormalized) spectral function
      $A_N(\omega)$ for $N=3$ (left hand side) and $N=4$ (right hand
      side) with respect to the ground-state energies $E_0^{2,7}$ and
      $E_0^{3,12}$, respectively. The full, dashed, and dotted lines
      correspond to different maximal angular momentum.}
\label{fig4}
\end{center}
\end{figure}

\section{Summary and Outlook}
We have presented a simple model of a graphene quantum dot, suitable for the
analysis of interaction effects. We show that Coulomb blockade effects are
similar to those in other systems when the chemical potential is far from the
neutrality point.

The Dirac equation which describes the electronic states of graphene allows
for the existence of midgap states, near defects or surfaces. The presence of
these states changes qualitatively the properties of the dot in the Coulomb
blockade regime. As the kinetic energy of these states is nearly zero, the
resulting wave function is mostly determined by the interaction, and deviates
significantly from a single Slater determinant. In order to describe
correlations beyond mean field we employed the method of configuration
interaction within the subspace of surface states. Since it is known that
screening is weak in the described case close to half filling, we considered
the electrons to interact with each other via the long-ranged Coulomb
interaction in contrast to a point-like Hubbard interaction studied for
example in Ref.\cite{Rossier07}.


Making use of the simple analytical form of the surface states, we have
identified two possible correlated wave functions which are in good agreement
with few-particle exact calculations: a wave function similar to that proposed
by Laughlin for the Fractional Quantum Hall effect, and another describing a
Wigner crystal.  These results indicate the existence of strong correlations,
although they do not allow us to analyze the existence of an incompressible
electron liquid in the thermodynamic limit.  We note that the correlations
present in the spin polarized states studied here, arise only for long-ranged
interactions, while they are absent if an effective point-like Hubbard
interaction is considered.  We expect the few particle states to be pinned by
disorder at the edges (note, however, that their extension can be comparable
to the dot size, so that the pinning will not be large).

The transport properties in the regime where the midgap states are partially
occupied will deviate from that observed in other quantum dots. The strongly
correlated nature of the wave function implies that non shake up effects will
suppress the tunneling density of states.

An interesting extension of this work concerns the valley- and spin-degree of
freedom which will be addressed in a subsequent publication.

\section{Acknowledgments}
We appreciate helpful discussions with A. H. Castro Neto, L. Brey,
A. F. Morpurgo,
K. S. Novoselov, and
F. Sols. This work has been supported by the EU Marie Curie RTN Programme
No.
MRTN-CT-2003-504574, the EU Contract 12881 (NEST), and by MEC
(Spain) through Grants No. MAT2002-0495-C02-01,
FIS2004-06490-C03-00, FIS2004-05120, FIS2005-05478-C02-01,  the Comunidad de
Madrid program CITECNOMIK, and the Juan de la
Cierva Programme.
\appendix
\section{Discretization of the Dirac equation}
\label{App:TightBinding}
The Dirac equation for angular momentum $l$ can be written as two coupled one
dimensional differential equations:
\begin{eqnarray}
V ( r ) \psi_A ( r ) + \vf \left( i \partial_r \pm i \frac{l+1}{r}
\right) \psi_B ( r ) &= &\epsilon \psi_A ( r ) \nonumber \\
\vf \left( i \partial_r \mp i \frac{l}{r} \right) \psi_A ( r ) + V (
r ) \psi_B ( r ) &= &\epsilon \psi_B ( r ) \nonumber \\ & &
\label{app_dirac}
\end{eqnarray}
where the two signs correspond to the two Dirac points. We now
analyze a given Dirac equation. Extension to the other valley is
straightforward. Equation (\ref{app_dirac}) can be written as:
\begin{widetext}
\begin{eqnarray}
V ( r ) \left[ \psi_A ( r ) + \psi_B ( r ) \right] + \vf  \left( i
\partial_r + \frac{i}{2 r} \right) \left[ \psi_A ( r ) +
\psi_B ( r ) \right] - i \vf \frac{2 l + 1}{2 r} \left[ \psi_A ( r )
- \psi_B ( r ) \right] &= &\epsilon \left[ \psi_A ( r ) + \psi_B ( r
)
\right] \nonumber \\
V ( r ) \left[ \psi_A ( r ) - \psi_B ( r ) \right] - \vf  \left( i
\partial_r + \frac{i}{2 r} \right) \left[ \psi_A ( r ) -
\psi_B ( r ) \right] + i \vf \frac{2 l + 1}{2 r} \left[ \psi_A ( r )
+ \psi_B ( r ) \right] &= & \epsilon \left[ \psi_A ( r ) - \psi_B (
r ) \right]
\end{eqnarray}
\end{widetext}
We define
\begin{eqnarray}
\tilde{\psi}_1 ( r ) &= &\frac{\psi_A ( r ) + \psi_B ( r
)}{\sqrt{r}}
\nonumber \\
\tilde{\psi}_2 ( r ) &= &\frac{\psi_A ( r ) - \psi_B ( r
)}{\sqrt{r}}
\end{eqnarray}
and we obtain:
\begin{eqnarray}
V ( r ) \tilde{\psi}_1 ( r ) + i \vf \partial_r \tilde{\psi}_1 ( r )
-i \vf \frac{2 l + 1}{2 r} \tilde{\psi}_2 ( r ) &= &\epsilon
\tilde{\psi}_1
( r ) \nonumber \\
V ( r ) \tilde{\psi}_2 ( r ) - i \vf \partial_r \tilde{\psi}_2 ( r )
+ i \vf \frac{2 l + 1}{2 r} \tilde{\psi}_1 ( r ) &= &\epsilon
\tilde{\psi}_2 ( r ) \nonumber \\ & & \label{Dirac}
\end{eqnarray}

A set of discrete equations which, taking the continuum limit, lead to
Eq.(\ref{Dirac}) is:
\begin{eqnarray}
\left( 1 - \frac{2l+1}{4 n} \right) a_n^l + \left( 1 + \frac{2l+1}{4
n} \right)
a_{n+1}^l + v_n b_n^l &= &\epsilon b_n^l \nonumber \\
\left( 1 + \frac{2l+1}{4 n} \right) b_{n-1}^l - \left( 1 -
\frac{2l+1}{4 n} \right) b_n^l + v_n a_n^l &= &\epsilon a_n^l  \nonumber \\
& &
\end{eqnarray}

This set of equations is formally equivalent to a dimerized tight binding
chain, as schematically shown in Fig.~\ref{hoppings}. These chains admit
zero energy estates localized at the ends, when the last hopping is smaller
than the previous one. In order to avoid the formation of a spurious level at
the center of the dot, $n=1$, the chain is doubled, as also shown in
Fig.~\ref{hoppings}.
\begin{figure}[t]
  \begin{center}
    \includegraphics*[width=5cm,angle=0]{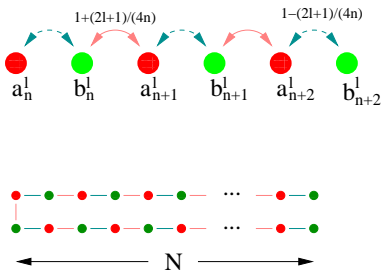}
    \caption{(color online) Top: Sketch of the discretization of the Dirac
      equation
    in radial coordinates used in the text. Bottom: doubled chain used in the
    calculations, in order to avoid spurious effects at $n=1$.}
\label{hoppings}
\end{center}
\end{figure}
The Coulomb potential is discretized as:
\begin{equation}
v_n =\sum_{m=1}^N v_{nm}  \sum_l \frac{{a_m^l}^2 + {b_m^l}^2}{m}
\end{equation}
and:
\begin{equation}
v_{nm} = v_0 \int_0^{2 \pi} d \theta \frac{m}{\sqrt{m^2 + n^2 + 2 m n \cos (
    \theta )}}
\end{equation}
In terms of the original Dirac equation, the energies are expressed in units
of $\hbar \vf/R$ and the parameter $v_0$ is given by $v_0
= (e^2 /4\pi \epsilon_0\epsilon)/ \hbar v_F\approx 1$.

\section{Kinetic Energy due to Next-Nearest Neighbor Hopping}
\label{App:KineticEnergy}
Due to next-nearest neighbor hopping $t'\sim0.1t$, the initially flat band of
surface states becomes dispersive. From Ref. \cite{PGN06}, the kinetic term
due to $t'$ is given by $T=\frac{9}{4}t'a^2 p^2\rightarrow -\frac{9}{4}
t'a^2\Delta$.
As for the Coulomb interaction we restrict our Hilbert space to the surface
states $\psi_m(r,\theta) =\Psi_{s=+}^B ( r , \theta )$ defined in Eq.~(\ref{edge_state}):
\begin{align*}
\langle m|T|n \rangle&=-\delta_{n m}\frac{9}{4}t'a^2 \int d^2 r \psi_m^*(\vec{r}) \Delta  \psi_n(\vec{r})\\
\int d^2 r \psi_m^* \Delta  \psi_m&=-\int d^2 r \nabla
\psi_m^* \nabla \psi_m+2\pi \left[\psi_m^* r \partial_r \psi_m\right]_{r=R}
\end{align*}
In the second row we used partial integration leading to the boundary
term (second term on right hand side). Including next nearest neighbor
hopping this boundary term has to vanish, while the general form of
the wavefunction is assumed to change only close to the boundary. We
thus only keep the first term which results in
\begin{eqnarray*}
\langle m|T|n \rangle=\delta_{n m}\frac{9 a^2 t'}{2 R^2}  m (m+1)=\frac{\hbar
  v_F}{R}\frac{3 a}{10 R}  m (m+1)\;.
 \end{eqnarray*}

\bibliography{qdotPaco}

\begin{thebibliography}{35}
\bibitem[{cha()}]{charge}
\bibinfo{note}{Note, however, that electrostatic considerations in two
  dimensions lead to a charge distribution which varies smoothly over the dot,
  but has a $\rho ( r ) \propto 1/\sqrt{R - r}$ divergence at the edges.}
\end{thebibliography}
\end{document}